\documentclass[preprint,aps]{revtex4}
\usepackage{graphics}
\usepackage{graphicx}
\usepackage{dcolumn}
\usepackage{bm}

\begin{document}
\title{Normal reflection at superconductor - normal metal interfaces due to Fermi surface mismatch}

\author{Elina Tuuli$^{1-3}$ and Kurt Gloos$^{1,3}$}

\address{$^1$ Wihuri Physical Laboratory, Department of Physics and Astronomy, FIN-20014 University of Turku, Finland}
\address{$^2$ Graduate School of Materials Research (GSMR), FIN-20500 Turku, Finland}
\address{$^3$ Turku University Centre for Materials and Surfaces (MatSurf), FIN-20014 Turku, Finland}


\begin{abstract}
Electrons can be reflected at an interface between two metals because of a dielectric barrier or different properties of the Fermi surface. Andreev reflection allows to directly measure normal reflection when one of the metals is a superconductor. We have investigated normal reflection  at interfaces between non-magnetic normal metals and superconducting Nb ($T_c = 9.2\,$K) and Al ($T_c = 1.2\,$K). The distribution of the values of the relative strength of the interface barrier, $Z$, for a number of contacts of a specific metal combination shows a well-defined peak which can be attributed to Fermi surface mismatch. Our reflection coefficients are generally smaller than those predicted theoretically or those derived from proximity-effect studies of normal-superconductor bi-layers. 
\end{abstract}

\maketitle 

\section{Introduction}

For practical nanoscale applications, including spintronics, the quality of an interface matters. To develop new devices it is important to understand their properties. The normal reflection coefficient $\mathcal{R}$, or the complementary transmission coefficient $\tau = 1-\mathcal{R}$, is one of the parameters that characterizes an interface. It informs how large a fraction of the incident electrical current is not transported through an interface. Reflection or transmission at metal interfaces has mainly been investigated with the help of superconducting proximity effect measurements~\cite{Cirillo2004,Tesauro2005,Attanasio2006,Kushnir2009} and the current perpendicular to the plane magnetoresistance (CPP-MR) studies~\cite{Stiles2000,Xu2006,Sharma2007,Park2000} using thin films. Point-contact Andreev reflection spectroscopy offers a different approach to this problem and does not require thin-film samples. It is a differential method as it  compares the normal reflection of the two-particle electron - hole process to that of a single-particle one. Andreev reflection at a superconductor - normal metal interface is described by the BTK-model~\cite{BTK} with only three adjustable parameters: the energy gap of the superconductor $2\Delta$, the temperature $T$ which is measured independently, and the relative strength $Z$ of a $\delta$-function interface barrier. This simple model has been extended to take into account the effects of spin polarization~\cite{Strijkers,Woods} and the finite Cooper pair lifetime~\cite{Plecenik}.

It has been suggested~\cite{Blonder1983} that the BTK $Z$ parameter consists of two parts, $Z_b$ describing the strength of a real tunneling barrier at the contact and $Z_0$ the reflection caused by the mismatch of the different electronic structures on each side of the interface. The $Z$ parameter is related to the transmission coefficient $\tau = 1/(1+Z^2)$. Without tunneling barrier only Fermi surface mismatch contributes, providing a lower limit for the reflection coefficient of a specific metal - metal combination when a tunneling barrier is absent. We present here the results of our point contact measurements on the normal reflection at interfaces between superconducting Nb and Al in contact with non-magnetic noble metals Ag, Cu and Pt.

\section{Experimental details and results}

	Most of the measurements with Nb ($T_c = 9.2$ K) were carried out at $4.2$ K in liquid helium. A smaller number of interfaces with Nb and all contacts with Al ($T_c = 1.2$ K) were measured down to $0.1$ K in vacuum. The contacts were formed between two pieces of wire (0.25 mm diameter) using the shear method. Their differential resistance was recorded in the standard four-wire scheme with low-frequency current modulation. A review of experimental point-contact techniques is given in~\cite{kirja}.
	
\begin{figure}
	\includegraphics[width=16cm]{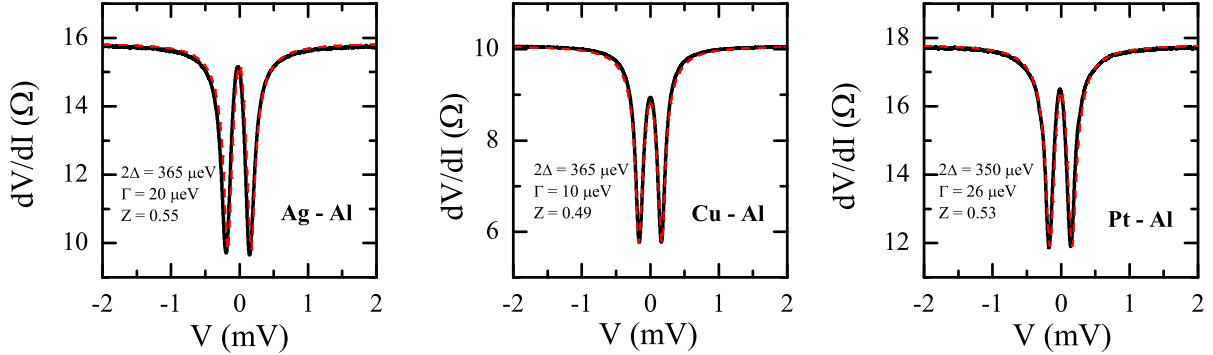}
	\caption{\label{spektrit}Typical Al - normal metal spectra measured at 0.1 K in vacuum. 
            The black solid line is the measured spectrum and the red dashed line is the fit from the modified BTK model.}
	\end{figure}

	The differential resistance spectra can be divided into different groups~\cite{ZBA,oma} of which we have analysed only those contacts with an Andreev-type spectra with or without side peaks using the modified BTK model that takes into account the finite Cooper pair lifetime~\cite{Plecenik}. The normal-state resistance was restricted to 1-100 $\Omega$. Figure~\ref{spektrit} displays representative spectra and their fits of normal metals in contacts with Al. Spectra with Nb look qualitatively similar~\cite{oma}.
	
\begin{figure}
\includegraphics[width=16cm]{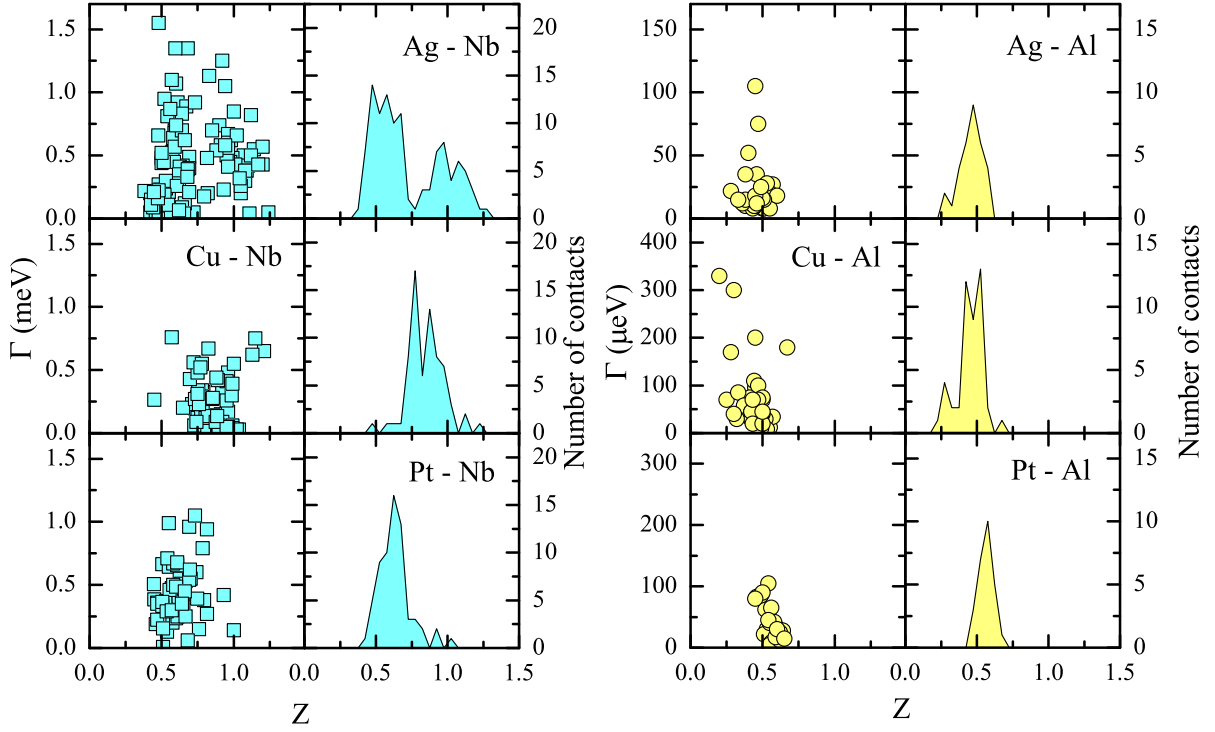}
\caption{\label{Nb}The distribution of the lifetime parameter $\Gamma$ and number of contacts of Nb - normal metal and Al - normal metal contacts as function of the strength of the interface barrier $Z$.}
\end{figure}

\begin{figure}
\includegraphics[width=16cm]{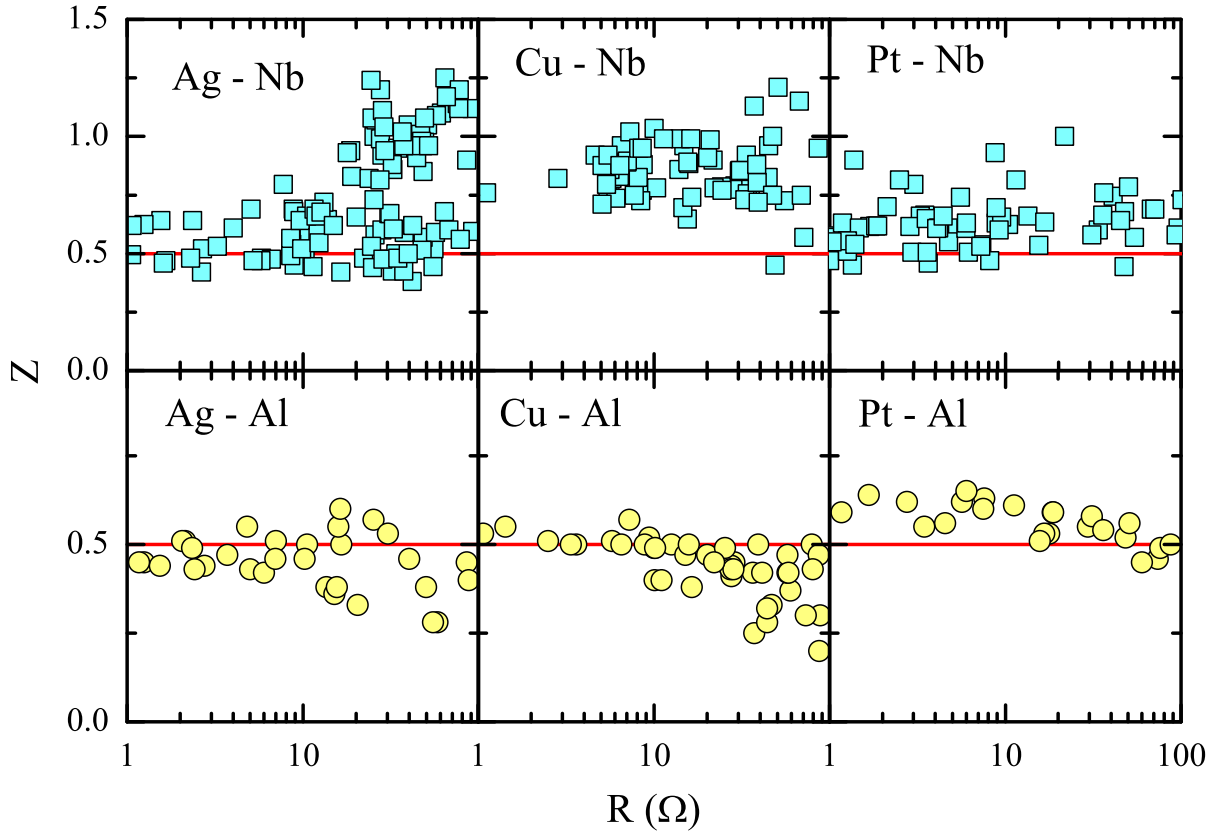}
\caption{\label{R}The $Z$ parameter as function of the contact resistance for Nb and Al in contact with normal metals. The thin red $Z = 0.5$ lines are a guide to the eye.}
\end{figure}

	Figure~\ref{Nb} summarizes the results of this analysis for Nb and Al contacts. The data points are strongly concentrated at preferred $Z$ values as can easily be seen in the histograms. The distributions of the contacts both with Nb and Al have clear onsets, finite widths and in some cases a weak tail at large $Z$ values. The Ag - Nb distribution consists apparently of two separate peaks. This could be caused by different preferred crystal orientations. The $Z$ values vary only weakly with the normal-state resistance $R$ in the specified range (Figure~\ref{R}).
	
\begin{table}
\caption{\label{taulukko} The $Z$ values for the investigated metal interfaces. The values marked with a * have been calculated from numerical data based on the free electron model in~\cite{Ashcroft} and ** are calculated from $\tau$ values in~\cite{Tesauro2005}.}
\begin{center}
\begin{tabular}{|c|c|c|c|c|c|c|}

\hline
 & Nb - Ag & Nb - Cu & Nb - Pt & Al - Ag & Al - Cu & Al - Pt\\
\hline
 Z (Onset) & 0.36 & 0.53 & 0.41 & 0.22 & 0.18 & 0.28 \\
 Z (Centre) & 0.57 & 0.82 & 0.62 & 0.48 & 0.40 &  0.57 \\
Z (Theory)* & 0.007 & 0.068 & & 0.19 & 0.13 & \\
Z (PE)** & 1.42 & 1.53 & & & &\\
\hline

\end{tabular}
\end{center}
\end{table}

\section{Discussion}

Interfaces between different metals should always have some amount of normal reflection caused by their different electronic structures, the so called Fermi surface mismatch (FSM). It is difficult to predict exactly the contribution of FSM to normal reflection because it depends strongly on the material and the crystal orientations at the interface. 

The onset of the $\Gamma(Z)$ histograms defines the smallest possible value for the $Z$ parameter (or the maximum transmission coefficient) which has been reached in our experiments. We believe that these contacts have a negligible tunneling barrier and the crystal directions are ``ideal''. The position of the peak agrees rather well with the average $Z$ value. There are also few data points at larger $Z$ values. These can be attributed to an additional weak tunneling barrier at the contact. The width of the $Z$ distribution is unlikely to be caused by tunneling because that should be less reproducible and lead to a wider variation of the observed $Z$ values. The $Z$ values obtained from our analysis are presented in Table~\ref{taulukko}. 

Thin film proximity effect measurements have lead to transmission coefficients of only approximately 0.3 for Nb in contact with non-magnetic normal metals~\cite{Tesauro2005}. CPP-MR studies result in transmission coefficients of similar size even though there should be no oxide barriers in either of these experiments~\cite{Tesauro2005,Sharma2007,Park2000}. We can only speculate that our much larger transmission coefficients could be caused, for example, by effects related to the different size of the samples or different properties of thin films and bulk wires. It might also be possible that the different measurement techniques observe different effects - we are investigating Andreev reflection as function of bias voltage and in the proximity effect as well as in the CPP-MR studies the transport is observed at zero bias. 

The histograms in Figure~\ref{Nb} also highlight the differences and similarities of the investigated interfaces. Apart from the clearly double-peaked Ag - Nb distribution, the other superconductor - normal metal histograms look surprisingly similar. All the normal metals have almost the same $Z$ with both Nb and Al despite their strongly different Fermi surfaces~\cite{netti}. 
On the other hand, the lifetime parameter $\Gamma$ at the interface does not depend on the normal metal, but differs for the two investigated superconductors. It appears to scale with the critical temperature $T_c$ or the superconducting energy gap $2 \Delta$. Both observations pose interesting questions for future research.

\section{Conclusions}

We have investigated normal reflection at interfaces between noble metals and superconducting Nb and Al. The measured differential resistance spectra were analysed with a modified BTK model which yielded a rather well-defined $Z$ parameter of normal reflection which seems to represent Fermi surface mismatch.

\section{Acknowledgements}
We thank the Jenny and Antti Wihuri Foundation and the Finnish Concordia Fund for financial support.


\section*{References}
\begin{thebibliography}{9}

\bibitem{Cirillo2004}
  Cirillo C, Prischepa S L, Salvato M and Attanasio C 2004
  {\it Eur. Phys. J.} B {\bf 38} 59

\bibitem{Tesauro2005}
  Tesauro A, Aurigemma A, Cirillo C, Prischepa S L, Salvato M and  Attanasio C 2005
  {\it Supercond. Sci. Technol.} {\bf 18} 1

\bibitem{Attanasio2006}
  Attanasio C 2006
  {\it Nanoscale Devices - Fundamentals and  Applications}
  ed Gross R {\it et al.}
   (Springer) p 241
  
\bibitem{Kushnir2009}
  Kushnir V N, Prischepa S L, Cirillo C and Attanasio C 2009
  \textit{J. Appl. Phys.} \textbf{106} 113917
  
\bibitem{Stiles2000}
  Stiles M D and Penn D R 2000
  \textit{Phys. Rev.} B \textbf{61} 3200
  
\bibitem{Xu2006}
  Xu P X, Xia K, Zwierzycki M, Talanana M and Kelly P J 2006
  \textit{Phys. Rev. Lett.} \textbf{96} 176602
  
\bibitem{Sharma2007}
  Sharma A, Romero J A, Theodoropoulou N, Loloee R, Pratt Jr. W P and  Bass J 2007
  \textit{J. Appl. Phys.} \textbf{102} 113916
  
\bibitem{Park2000}
  Park W, Baxter D V, Steenwyk S, Moraru I, Pratt Jr. W P and Bass J 2000
  \textit{Phys. Rev.} B \textbf{62} 1178
  
\bibitem{BTK} Blonder G E, Tinkham M and Klapwijk T M 1982 {\it Phys. Rev.} B {\bf 25} 4515

\bibitem{Strijkers} Strijkers G J, Ji Y, Yang F Y and Chien C L 2001 {\it Phys. Rev.} B {\bf 63} 104510

\bibitem{Woods} Woods G T, Soulen Jr. R J, Mazin I, Nadgorny B, Osofsky M S, Sanders J, Srikanth H, Egelhoff W F and Datla R 2004 {\it Phys. Rev.} B {\bf 70} 054416

\bibitem{Plecenik} Plecen\'{i}k A, Grajcar M and Be\v{n}a\v{c}ka \v{S} 1994 {\it Phys. Rev.} B {\bf 49} 10016

\bibitem{Blonder1983} Blonder G E and Tinkham M 1983 {\it Phys. Rev.} B {\bf 27} 112

\bibitem{kirja} Naidyuk Yu G and Yanson I K 2005 {\it Point-Contact Spectroscopy} (Springer)

\bibitem{ZBA} Gloos K 2009 {\it Fiz. Nizk. Temp.} {\bf 35} 1204

\bibitem{oma} Tuuli E and Gloos K 2011 {\it Fiz. Nizk. Temp.} {\bf 37} 609

\bibitem{Ashcroft} Ashcroft N W and Mermin N D 1976 {\it Solid State Physics} (Brooks/Cole)

\bibitem{netti} http://www.phys.ufl.edu/fermisurface/


\end{thebibliography}
\end{document}